\documentclass[aps,preprint,showpacs,amsfonts]{revtex4-1}

\usepackage{amsmath}
\usepackage{latexsym}
\usepackage{float}
 \usepackage{amssymb}
\usepackage{graphicx}
\usepackage{textcomp}
\usepackage{hyperref}
\usepackage{subfigure}

\def\ba{\begin{eqnarray}}
\def\ea{\end{eqnarray}}
\def\be{\begin{equation}}
\def\ee{\end{equation}}
\def\bm{\begin{math}}
\def\me{\end{math}}

\newcommand{\dummy}

\begin{document}

\title{Coarsening in 3D Nonconserved Ising Model at Zero Temperature: 
Anomalies in structure and relaxation of order-parameter autocorrelation}
\author{Saikat Chakraborty and Subir K. Das$^{*}$}
\affiliation{Theoretical Sciences Unit, Jawaharlal Nehru Centre for Advanced Scientific Research,
Jakkur P.O, Bangalore 560064, India
}

\date{\today}

\begin{abstract}
Via Monte Carlo simulations we study pattern and aging during coarsening in nonconserved nearest neighbor Ising model, 
following quenches from infinite to zero temperature, in space dimension $d=3$. The decay 
of the order-parameter autocorrelation function is observed to obey a power-law behavior in 
the long time limit. However, the exponent of the power-law, estimated accurately via a 
state-of-art method, violates a well-known lower bound. This surprising fact has been 
discussed in connection with a quantitative picture of the structural anomaly that the 3D Ising model 
exhibits during coarsening at zero temperature. These results are compared with those for 
quenches to a temperature above that of the roughening transition.
\end{abstract}
\maketitle
\section{Introduction}
Kinetics of phase transitions \cite{on,bra}, following quenches of homogeneous systems to the state points 
inside the coexistence regions, remains an active area of research \cite{dfish,liu,yeu,hen,lor,%
mid,mid2,maj,zan,oh,por,oon,yeu2,abr,das,sho,cor,ole,ole2,cha,sat,der,sah,bla,cha2,maze}. 
A particular interest has been in the case \cite{on,bra} when the temperature 
($T$) of a magnetic system, prepared at the paramagnetic region, is suddenly lowered to a value that 
corresponds to the ferromagnetic region of the phase diagram. Following such a quench 
the system evolves towards the new equilibrium via formation 
and growth  of domains rich in atomic magnets aligned in a particular direction. For such an 
evolution, in addition to the understanding of time ($t$)-dependence of the average domain size 
($\ell$) \cite{on,bra,sho,cor,ole,ole2,cha,all}, there has been significant interest in obtaining 
quantitative information on pattern formation \cite{oh,por,oon,yeu2,abr,das}, 
persistence \cite{sat,der,sah,bla,cha2} and aging \cite{dfish,liu,yeu,hen,lor,mid,mid2,maj,zan,maze}. 
\par
Ising model \cite{on,bra} has been instrumental in understanding of the above aspects 
of kinetics of phase transitions. Via computer simulations of this model a number of 
theoretical expectations have been confirmed \cite{bra}. Some of these we describe below in the 
context of nonconserved order parameter. 
\par
The (interfacial) curvature driven growth in this case is expected to provide \cite{bra,all}
\begin{equation}\label{e_growth}
\ell \sim t^{\alpha}; ~ \alpha=1/2,
\end{equation}
referred to as the Cahn-Allen growth law. The two-point equal-time 
correlation function \cite{bra}, that quantifies the pattern, in this context, was 
obtained by Ohta, Jasnow and Kawasaki (OJK) \cite{oh}, and has the form 
\begin{equation}\label{e_ojk}
C(r,t)=\frac{2}{\pi}\sin^{-1}\gamma,
\end{equation}
where
\begin{equation}\label{e_gamma}
\gamma=\exp(-r^{2}/8Dt),
\end{equation}
$D$ being a diffusion constant and $r$ the scalar distance between 
two space points $\vec{r}_{1}$ and $\vec{r}_2$. 
\par
Note that $C(r,t)$ is a special case of a more general two-point two-time 
(space and time-dependent) order-parameter ($\psi$) correlation function \cite{liu}
\begin{eqnarray}\label{e_corfn}
C_{\textrm{gen}}(\vec{r}_{1},\vec{r}_{2},t,t_{w})=\langle \psi(\vec{r}_{1},t)\psi(\vec{r}_{2},t_{w})\rangle-
\langle \psi(\vec{r}_{1},t)\rangle \langle\psi(\vec{r}_{2},t_{w})\rangle,
\end{eqnarray}
when $t=t_w$ and the pattern is isotropic. On the other hand, when $\vec{r}_{1}=\vec{r}_{2}$, 
$C_{\textrm{gen}}$ is referred to as the two-time autocorrelation function \cite{dfish,liu}. This we will denote by 
$C_{\textrm{ag}}(t,t_{w})$, where $t$ and $t_w$ ($\geqslant t$) are referred to as the 
observation and waiting times, respectively. For $C_{\textrm{ag}}$ there exists prediction 
of power-law decay as \cite{dfish,liu}
\begin{equation}\label{e_agscl}
C_{\textrm{ag}} \sim \bigg (\frac{\ell}{\ell_{w}}\bigg)^{-\lambda},
\end{equation}
where $\ell_{w}$ is the average domain size at time $t_{w}$. For the aging exponent $\lambda$, 
Fisher and Huse (FH) \cite{dfish} predicted a lower bound 
\begin{equation}\label{e_fh}
\lambda \geqslant \frac{d}{2},
\end{equation}
where $d$ is the space dimension.
\par 
Monte Carlo (MC) simulations \cite{lan} of the Ising model have been performed \cite{bra,das}
in various space dimensions, for quenches to various values of $T$. In $d=2$ the above predictions 
were found to be valid, irrespective of the temperature of quench. 
The status is similar with respect to simulations in $d=3$ at reasonably high 
temperatures. On the other hand, from the limited number of available works, 
it appears that the coarsening of the 3D Ising model at $T=0$ is 
special \cite{sho,cor,ole,ole2,cha,cha2}. Following reports 
on the slower growth and unusual structure in 
this case, we have undertaken comprehensive study of aging phenomena 
via the calculations of $C_{\textrm{ag}}(t,t_{w})$, alongside obtaining a quantitative 
picture of the structural anomaly.
\par
We have obtained the scaling property of $C_{\textrm{ag}}(t,t_{w})$ and quantified its functional 
form via analysis of results from extensive MC simulations of very large systems. We observe 
scaling with respect to \cite{dfish} $x$ ($=\ell/\ell_w$) and power-law decay in 
the asymptotic limit. The correction to this power-law, in the 
small $x$ region, resembles that 
of the high temperature quench \cite{mid}. The exponent of the power-law has been 
estimated via the calculation and convergence of an appropriate 
instantaneous exponent  \cite{hus} in the asymptotic limit ($x\rightarrow \infty$).  
The value, thus extracted, surprisingly, violates the FH lower bound \cite{dfish}. This 
striking fact we have discussed in connection with the structural property \cite{yeu}. 
Preliminary results on this issue were reported in Ref. \cite{das}. However, in this 
earlier work violation of the FH bound was not observed. 
\section{Methods}
We implement the nonconserved dynamics \cite{bra} in the MC simulations of the Ising model by using 
spin-flips \cite{bra,lan,gla} as the trial moves. Essentially, we have randomly chosen a spin and changed 
its sign. The energies before and after a trial was calculated from the 
Ising hamiltonian \cite{on,bra,dfish} ($<ij>$ in the summation represents nearest neighbors)
\begin{equation}\label{e_ising}
H=-J \sum_{<ij>} {S_{i}S_{j}, S_{i}=\pm 1, J>0}.
\end{equation}
Following this, the moves were accepted in accordance with the standard 
Metropolis algorithm \cite{dfish}, based on the difference in energies between the original and 
the perturbed configurations. One MC step (MCS), the time unit used in 
our simulations, consists of $N$ trial moves, where 
$N$ is the total number of spins in the system. We have considered periodic boxes of 
simple cubic type such that $N=L^3$, where $L$ is the linear dimension of a cubic 
box, in units of the lattice constant.
\par
We present results from two different temperatures, viz., $T=0$ and $0.6T_c$, $T_c$, the critical temperature,  
being equal to $4.51 J/k_{B}$, where $k_{B}$ is the Boltzmann constant. 
In the following we set $k_{B}$, interaction strength $J$ and the lattice constant to unity. 
For both the temperatures, we start with random initial 
configurations that mimic infinite temperature scenario. Note that $T=0.6T_c$ 
lies above the roughening transition temperature \cite{abr,cor,van}. All results are presented after 
averaging over a minimum of $20$ independent initial configurations. Here note that 
the spin variable $S_i$ is same as the order parameter $\psi$ that is used in the 
definition of the correlation function.
\par
For the calculation of $C(r,t)$ and $\ell$, thermal noise at $0.6T_{c}$ was 
eliminated via application of a majority spin rule \cite{maj2}. While $\ell$ 
can be calculated from the scaling property of $C(r,t)$ (see discussion in results part) as
\begin{equation}\label{e_a}
C(\ell,t)=a,
\end{equation}
in this 
work we have also obtained it from the first moment of the domain-size distribution 
function \cite{maj2}, $P(\ell_{d},t)$, as 
\begin{equation}\label{e_dist}
\ell=\int{\ell_{d} P(\ell_{d},t)} d\ell_{d},
\end{equation}
where $\ell_{d}$ is the distance between two consecutive domain boundaries along any 
Cartesian direction. In the exercise related to scaling property of $C(r,t)$ we will 
use $\ell$ obtained from Eq. (\ref{e_a}), by setting $a$ to $0.5$. On the other hand, 
for quantifying the aging property via $C_{\textrm{ag}}(t,t_{w})$, we will use $\ell$ 
calculated via Eq. (\ref{e_dist}). Note that there exist other methods as well, for the 
calculation of $\ell$. All these methods provide results proportional to each other.
\section{Results}
\begin{figure}[htb]
\centering
\includegraphics*[width=0.45\textwidth]{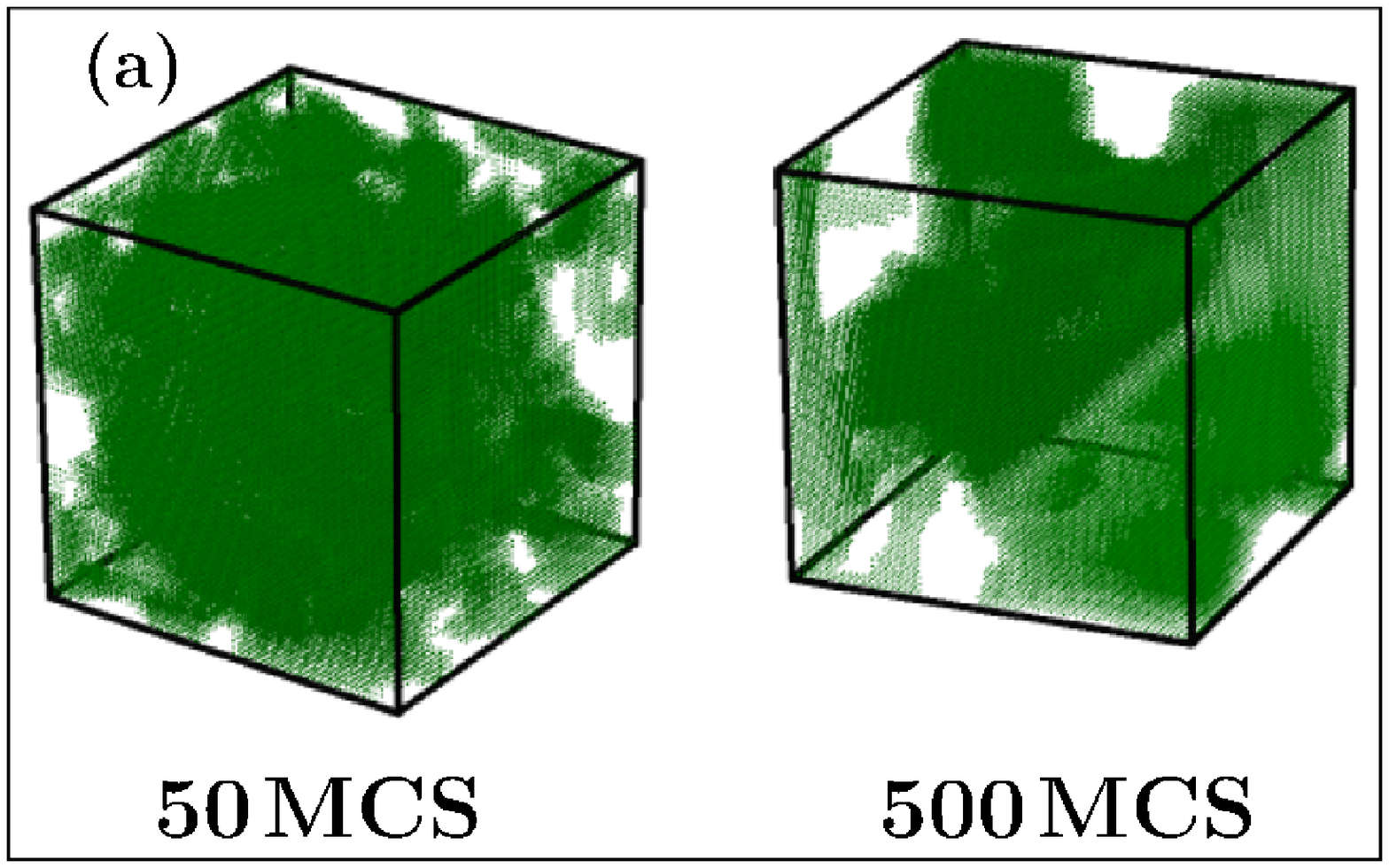}
\vskip 0.3cm
\includegraphics*[width=0.45\textwidth]{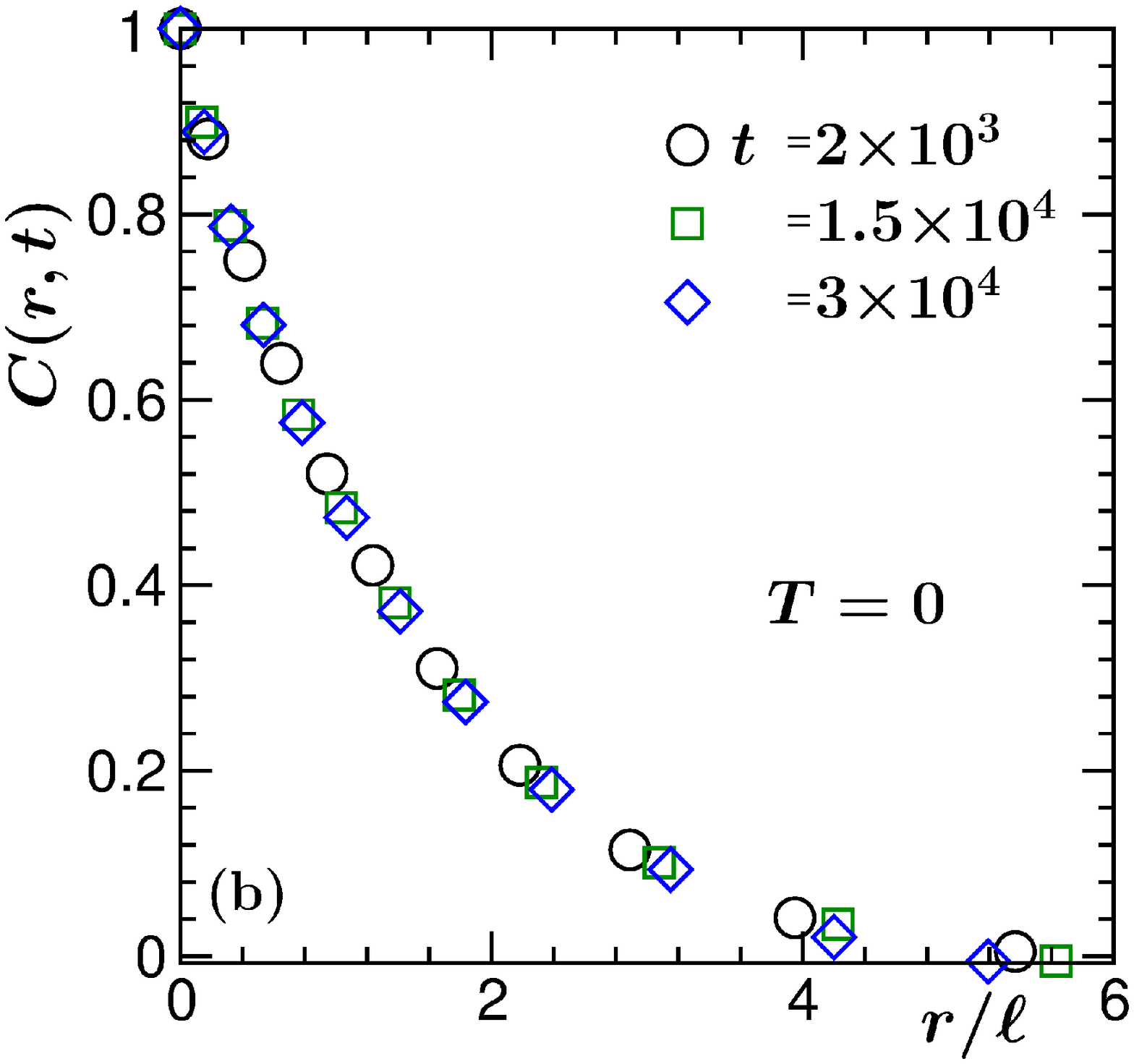}
\caption{\label{fig1} (a) Snapshots during the evolution of the 3D nonconserved Ising model. 
Pictures from two different times are shown. The locations of the ``up'' 
spins are marked. The linear dimension of the system is $L=64$.
(b) Plots of two-point equal-time correlation function versus $r/\ell$. Data from 
three different times are included. The box size corresponds to $L=512$. All results 
are from $T=0$.
}
\end{figure}
In Fig. \ref{fig1}(a) we show two snapshots during the evolution of the 3D Ising model, 
following a quench from infinite temperature to $T=0$. Growth in the system is clearly 
visible. To check for the structural self-similarity in the growth, 
in Fig. \ref{fig1}(b) we have plotted $C(r,t)$ versus $r/\ell$. 
Nice collapse of data from various different times imply the scaling property\cite{bra}:
\begin{equation}
C(r,t)=\tilde{C}(r/\ell(t)),
\end{equation}
where $\tilde{C}(y)$ is independent of time, requirement for the self-similarity. 
While this qualitative feature is the same as that exhibited \cite{bra,das} by the model 
in $d=2$ or for quenches to much higher values of $T$ in $d=3$, we will later see that $\tilde{C}$ here differs 
from that for the latter cases. This feature may have important consequence in the 
aging property. Unless otherwise mentioned, all results below are for quenches to $T=0$. 
Here note that the collapse of the $C(r,t)$ from $t <1000$ on the presented data sets 
in Fig. \ref{fig1}(b) is not as good.
\par
\begin{figure}[htb]
\centering
\includegraphics*[width=0.45\textwidth]{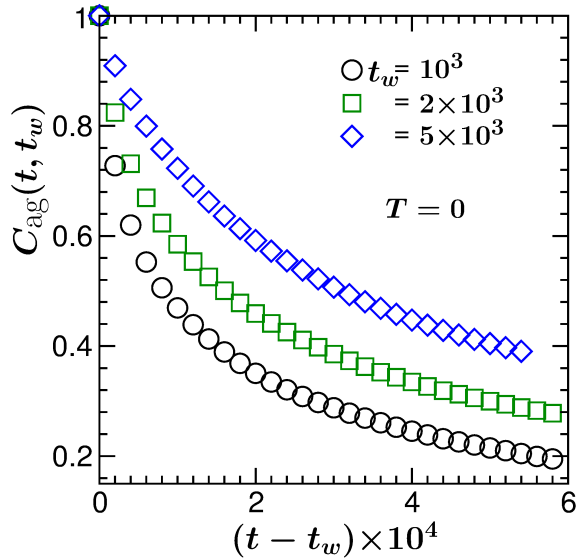}
\caption{\label{fig2} Plots of autocorrelation function versus $t-t_{w}$. Data from three different values of 
$t_{w}$ are shown. All results are for $T=0$ and $L=512$.
}
\end{figure}
\begin{figure}[htb]
\centering
\includegraphics*[width=0.45\textwidth]{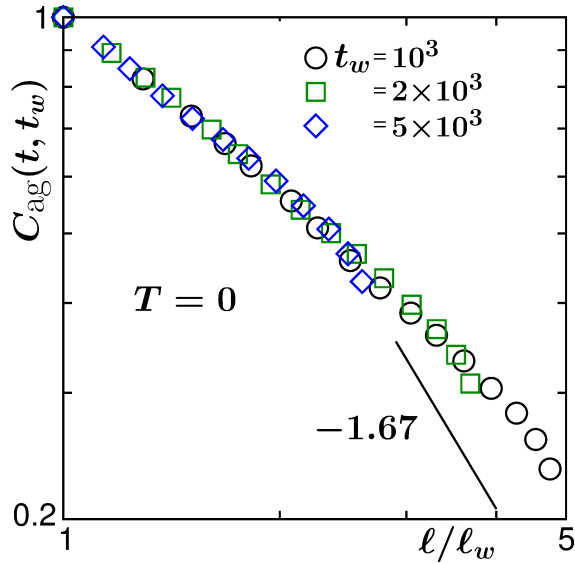}
\caption{\label{fig3} Scaling plot of $C_{\textrm{ag}}(t,t_{w})$ versus $\ell/\ell_{w}$, 
using data from three different values of $t_w$, on a log-log scale. The solid line 
represents a power-law, exponent being mentioned next to it. The presented results 
are from $T=0$ and $L=512$. 
}
\end{figure}

Next we focus our attention to the aging property.  
In Fig. \ref{fig2} we present plots of $C_{\textrm{ag}}(t,t_{w})$ 
versus $t-t_w$, from three different values of $t_w$. As expected, no time translation 
invariance is observed \cite{zan} and the decay becomes slower with the increase 
of $t_w$, implying aging in the system. However, the autocorrelation function for different $t_w$ values 
exhibit data-collapse when plotted versus $\ell/\ell_w$ \cite{dfish}. This is demonstrated in Fig. \ref{fig3}. 
The solid line in this figure represents a power-law with exponent $\lambda=1.67$. This value 
was predicted by Liu and Mazenko (LM) \cite{liu}, 
via a calculation that uses a Gaussian auxiliary field ansatz \cite{bra,liu}. 
LM constructed a dynamical equation for 
$C_{\textrm{gen}}(\vec{r}_{1},\vec{r}_{2},t,t_{w})$ as \cite{liu}
\begin{equation}\label{gen_cor}
\frac{\partial C_{\textrm{gen}}(\vec{R},t,t_{w})}{\partial t}={\nabla}^2 C_{\textrm{gen}}(\vec{R},t,t_{w})+
\frac{K}{t} C_{\textrm{gen}}(\vec{R},t,t_{w}),
\end{equation}
where $\vec{R}=\vec{r}_{1} - \vec{r}_{2}$ and the constant $K$ 
depends upon $d$. The (approximate) solution of this equation, in the 
asymptotic limit, for $\vec{R}=0$, provides a power-law for $C_{\textrm{ag}}(t,t_{w})$:
\begin{equation}
C_{\textrm{ag}}(t,t_{w}) \sim \ell^{-(d-2K)}.
\end{equation}
Given that \cite{liu} $K=0.6637$ in $d=3$, one obtains $\lambda=d-2K \simeq 1.67$.
\par
The LM line in Fig. \ref{fig3}, however, is in significant disagreement with the simulation data, 
even for very large value of $\ell$. Note that the presented data 
sets cover an overall length scale range $[50,250]$. However, as observed in previous studies in different 
dimension or at other temperatures, here also the scaling function exhibits continuous 
bending \cite{mid,maj}. Even though a fair agreement of the simulation data with the LM prediction is not 
yet observed, a trend for arriving at a better agreement with the latter or at least 
with a power-law behavior may be appreciated with the increase of $x=\ell/\ell_{w}$. 
This bending perhaps implies \cite{mid} presence of correction(s) for smaller $x$. 
In such a situation, if indeed a power-law behavior is expected in 
the $x \rightarrow \infty$ limit, to estimate the exponent
it is instructive to calculate the instantaneous exponent as \cite{mid,hus}
\begin{equation}
\lambda_{i}= -\frac{d \ln C_{\textrm{ag}}(t,t_{w})}{d \ln x}.
\end{equation}
\begin{figure}[htb]
\centering
\includegraphics*[width=0.45\textwidth]{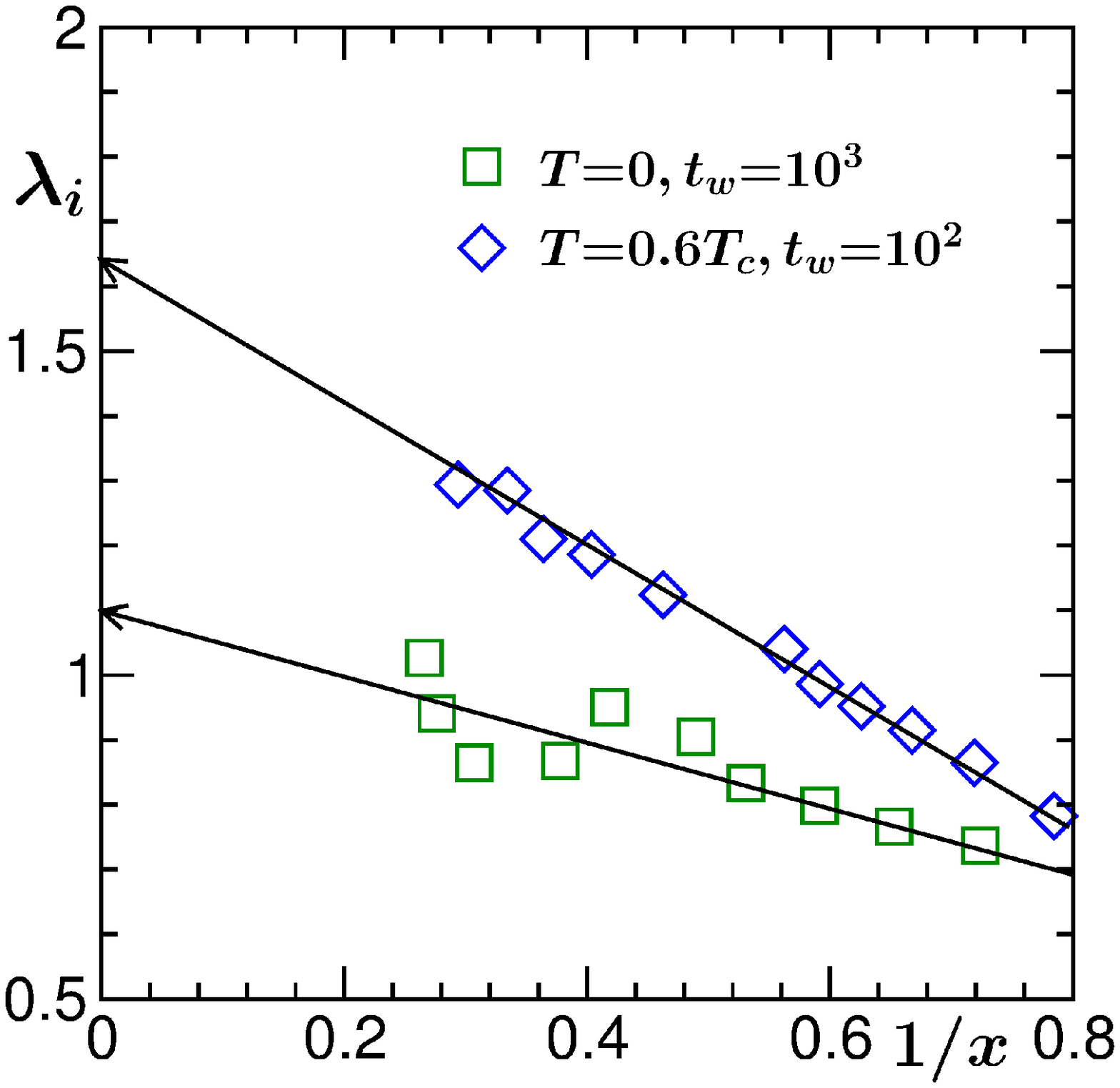}
\caption{\label{fig4} Plots of instantaneous exponent $\lambda_{i}$ versus $1/x$. We have presented data 
from $T=0$ and $0.6T_c$. The solid lines are guides to the eye. All results were obtained with $L=512$. 
We have done running averaging and thinned out data for the clarity of presentation. 
Oscillations in the $T=0$ data is due to strong fluctuations in each initial configurations.
}
\end{figure}
\par
We have calculated $\lambda_{i}$ for the scaling functions from $T=0$ and $T=0.6T_{c}$. 
These are plotted versus $1/x$ in Fig. \ref{fig4}. In both the cases linear convergence 
to the $x \rightarrow \infty$ limit is visible \cite{mid}. Such an extrapolation for the $T=0.6T_{c}$ 
data indeed leads to a number that is consistent with the LM \cite{liu} 
value $1.67$. Here note that in a later work \cite{cha2}, 
a modified value of $\lambda$, about $6\%$ smaller than $1.67$, was mentioned.  
On the other hand, for $T=0$ the convergence is to a 
much smaller value, $\simeq 1.1$. This number 
not only is significantly smaller than the LM \cite{dfish} value, it also violates the 
FH (lower) bound by a huge margin. Here note that the linear trend exhibited by 
the data sets in Fig. \ref{fig4} imply an exponential correction factor 
such that \cite{mid}
\begin{equation}
C_{\textrm{ag}}(t,t_{w})=Ae^{-\frac{B}{x}}x^{-\lambda},
\end{equation}
where $A$ and $B$ are constants.
\par
\begin{figure}[htb]
\centering
\includegraphics*[width=0.45\textwidth]{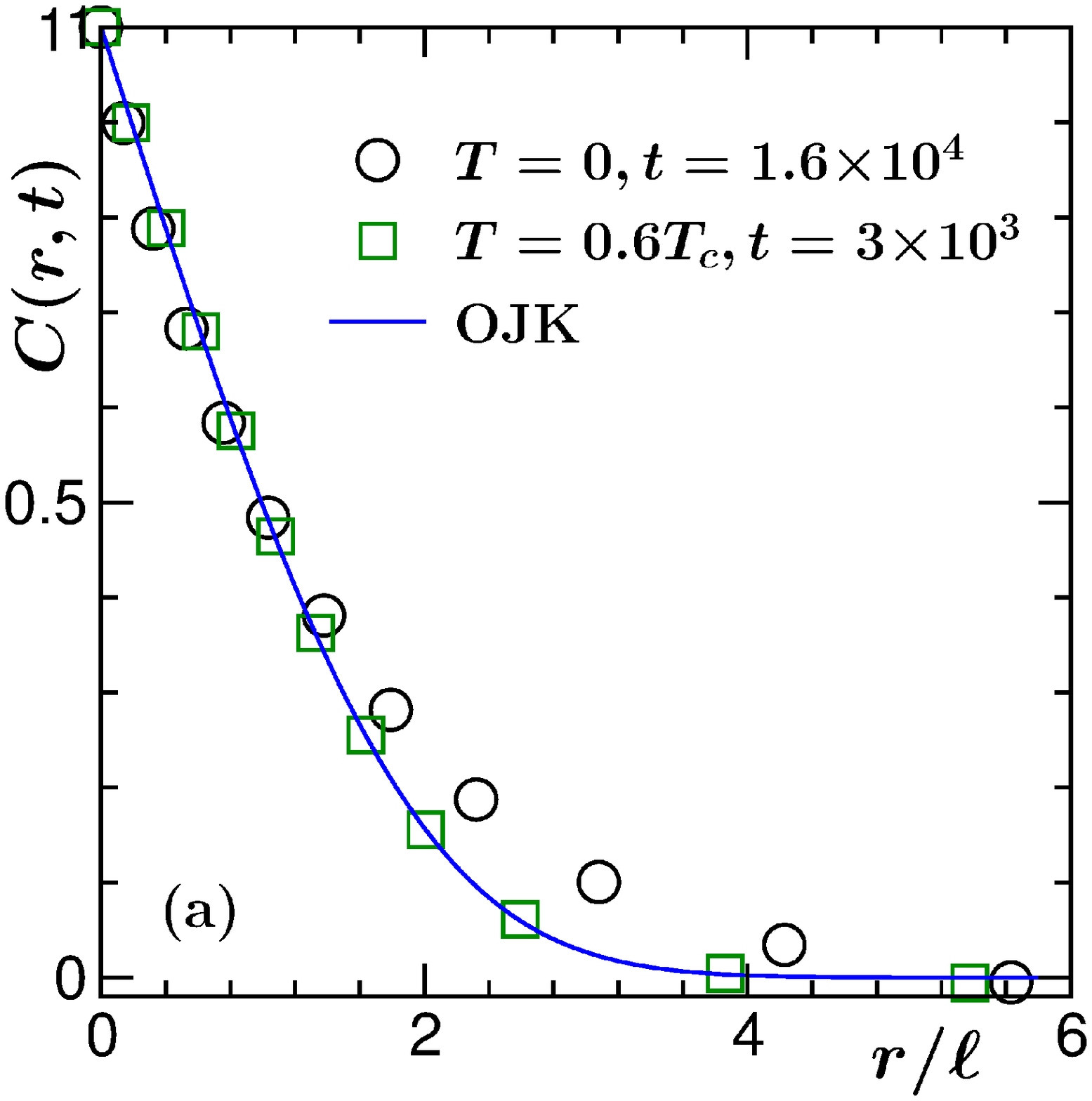}
\vskip 0.3cm
\includegraphics*[width=0.45\textwidth]{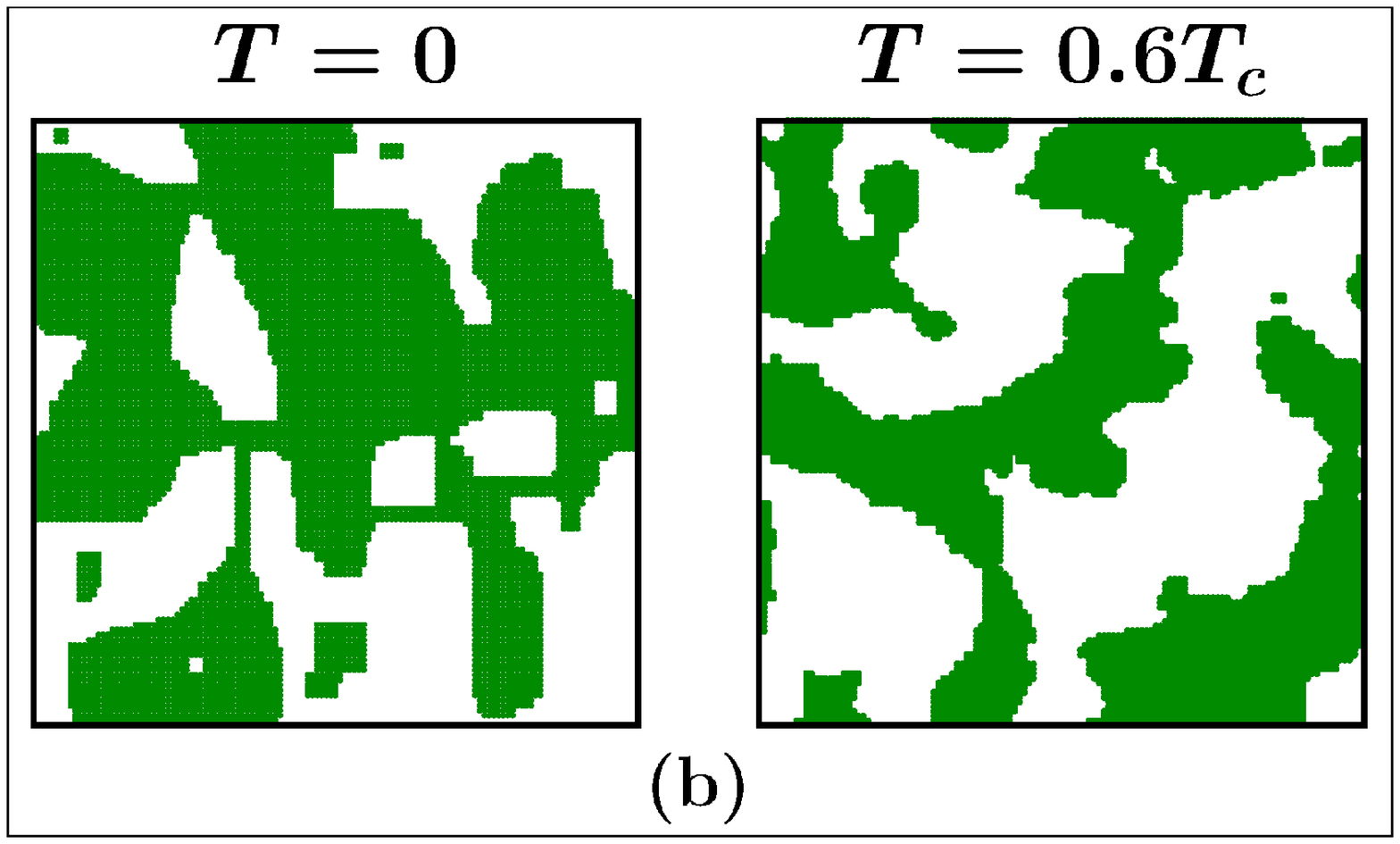}
\caption{\label{fig5} (a) A comparison of the two-point equal time correlation functions from $T=0$ and 
$0.6T_c$. In both the cases data from one time are presented, after 
scaling the distance axis by $\ell$. The continuous line there is the OJK analytical 
form. 
(b) Parts of 2D slices of snapshots from the evolutions at $T=0$ and $T=0.6T_c$. For $T=0.6T_c$, 
in the snapshot as well as for the calculation of $C(r,t)$ the thermal noise was 
removed (see the text).
}
\end{figure}

At this point we recall a recent observation of structural 
differences \cite{ole,ole2} of $T=0$ coarsening dynamics of 3D nonconserved Ising model with 
other situations. For that matter, in Fig. \ref{fig5}(a) we show a comparison of 
$C(r)$ at $T=0$ with that at $0.6T_c$. There exists significant difference 
between the two cases \cite{das}. The one at $0.6T_c$ is in nice agreement with the 
OJK function (see the continuous line).
\par
To understand the difference in the decay of $C_{\textrm{ag}}(t,t_{w})$ 
between the two chosen temperatures, we ask 
the question if the above mentioned structural mismatch is responsible for that. 
Here we note, Yeung, Rao and Desai (YRD) \cite{yeu} mentioned that the FH 
bound should be valid for only nonconserved order-parameter dynamics. 
The latter type of dynamics, of course, is being studied in this paper. 
The above point is raised \cite{yeu} by considering the known structural 
differences between conserved and nonconserved cases. Nevertheless, since, 
by now, we know that there exists difference between $T=0$ and higher temperature structures
even within the nonconserved framework \cite{das,ole,ole2}, further discussion and 
results with respect to this is worth presenting.
\par
\begin{figure}[h!]
\centering
\includegraphics*[width=0.45\textwidth]{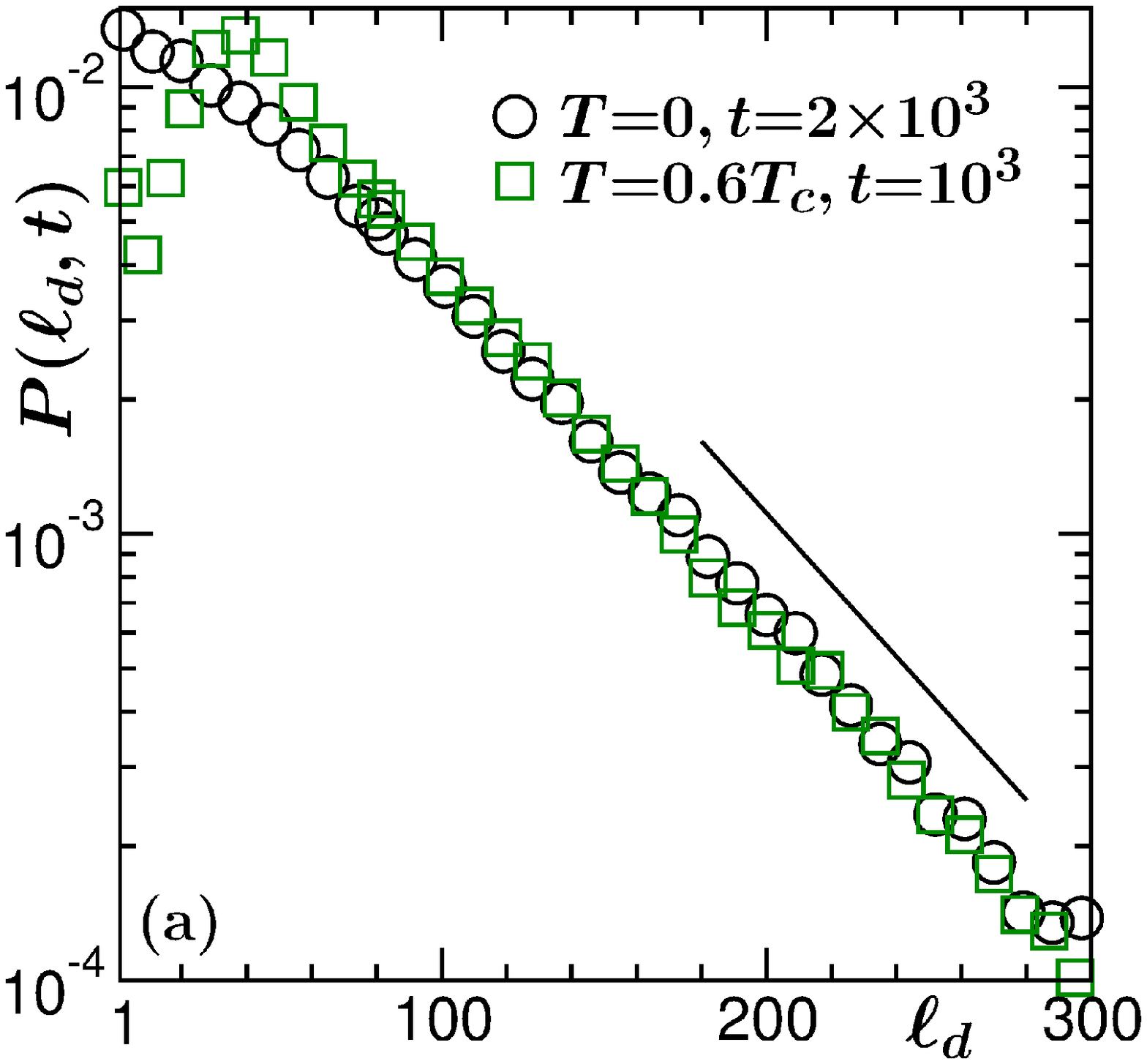}
\vskip 0.3cm
\includegraphics*[width=0.435\textwidth]{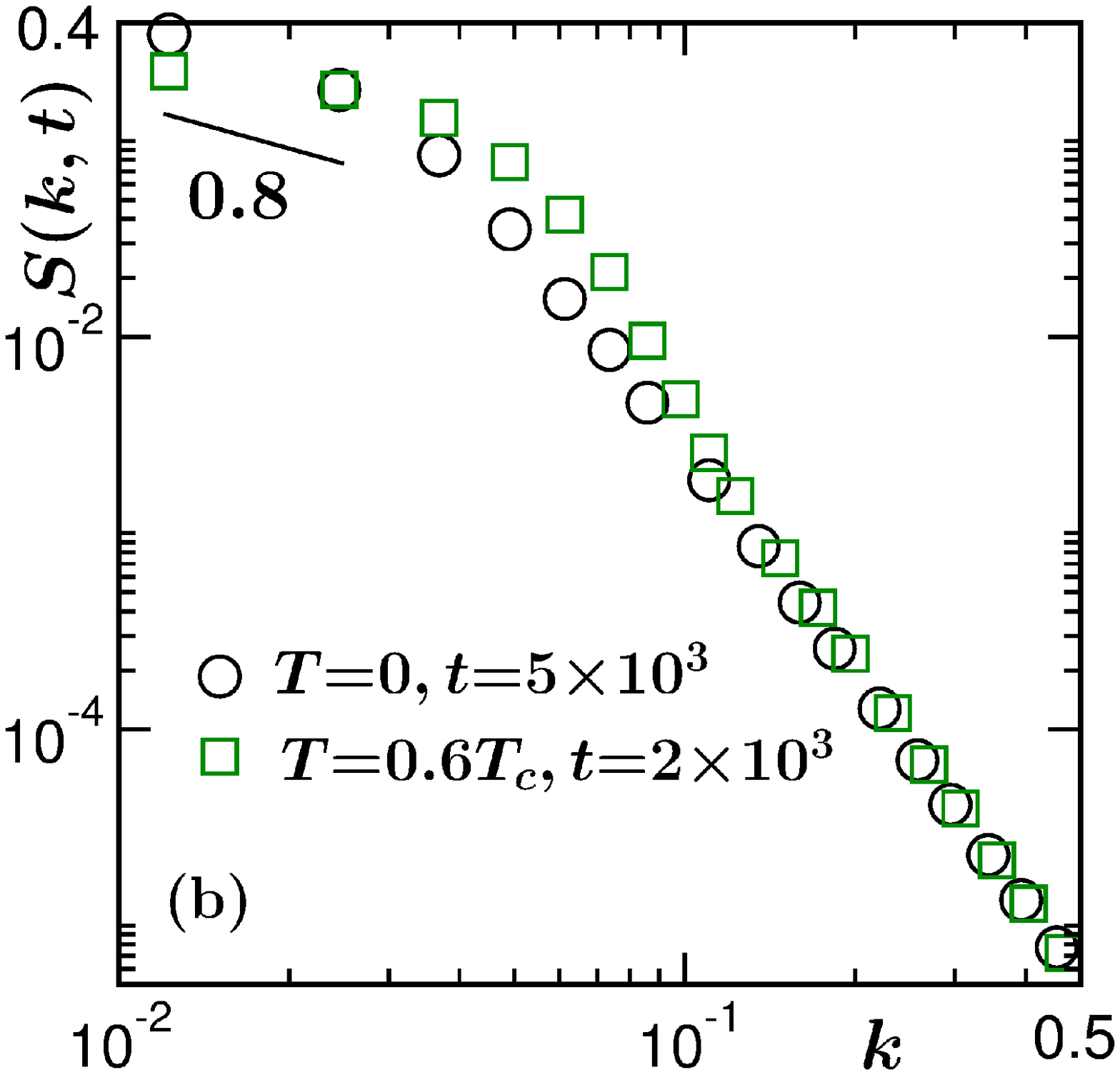}
\caption{\label{fig6} (a) Log-linear plots of the domain-size distribution function $P$ versus 
$\ell_d$. Results from $T=0$ and $0.6T_{c}$ are shown. The solid line represents an exponential decay. 
(b) Log-log plots of the equal-time structure factor vs $k$, at $T=0$ and $0.6T_{c}$. 
The solid line represents a power-law, exponent for which is mentioned.
All results correspond to $L=512$. The times are mentioned in the figures and are 
chosen in such a way that the characteristic length scales at the two temperatures 
are approximately same. 
}
\end{figure}

YRD obtained a modified lower bound \cite{yeu}
\begin{equation}\label{e_yrdbound}
\lambda \geqslant \frac{d+\beta}{2},
\end{equation}
where $\beta$ is the exponent for the small wave-vector ($k$) power-law behavior 
of structure factor [Fourier transform of $C(r,t)$] \cite{yeu2}:
\begin{equation}\label{e_smallk}
S(k,t) \sim k^{\beta}.
\end{equation}
In the case of usual nonconserved Ising dynamics $\beta=0$ and YRD 
bound coincides with the FH bound. Question now arises on the value 
of $\beta$ at $T=0$. Recall that there exists difference in $C(r,t)$ starting 
from intermediate length scale. This is consistent with the previous report 
that observed sponge-like structure \cite{ole,ole2}. We also find holes inside the domains 
of ``up'' and ``down'' spins. 
See the two-dimensional cuts of the snapshots in Fig. \ref{fig5}(b), obtained from the evolutions following 
quenches to $T=0$ and $0.6T_c$. Essentially the domains of the two types of spins 
are inter-penetrating in an unusual manner, at $T=0$, and creating 
a porous structure. In that case we expect a different form of 
the $P(\ell_{d},t)$ at $T=0$ from that at $T=0.6T_{c}$. This is shown in Fig. \ref{fig6}(a). 
Even though the large $\ell_{d}$ behavior is exponential for both the temperatures 
(see the log-linear choice of the plots), indeed 
the small $\ell_{d}$ behavior is quite different in the 
two cases, due to the presence of the porousity at $T=0$. 
This feature is consistent with the above mentioned difference in $C(r,t)$.  
Such difference in $C(r,t)$ is expected to provide 
disagreement in small $k$ behavior of $S(k,t)$ between the two temperatures.
\par
In Fig. \ref{fig6}(b) we present log-log plots of $S(k,t)$ versus $k$, for 
the two chosen temperatures. The small $k$ behavior is certainly not consistent with $\beta=0$. 
However, for $T=0.6T_c$, value of $\beta$ is very close to zero ($\simeq 0.2$).
(Note here that in $d=2$ or at high $T$ in $d=3$  
various authors \cite{yeu,yeu2,das} confirmed that $\beta \simeq 0$.) 
For the validity of the YRD bound at $T=0$ one requires $\beta \simeq -0.8$. The small $k$ 
behavior of $S(k,t)$ at $T=0$ is reasonably consistent with this number -- see the solid 
line in Fig. \ref{fig6}(b). We should mention here that for finite $L$ 
it is not possible to access very small values of $k$. If access of $k$ very close to zero  
becomes possible by considering large $L$, one may observe a behavior consistent 
with $\beta=0$ even for $T=0$, though such a convergence may be much slower than that for the 
higher temperature case. Furthermore, another question arises, what upper value of $k$ 
should be considered to be small. This can be answered by knowing the average 
domain size that grows with a slow rate at $T=0$.  
For this purpose we provide a brief discussion on how the YRD bound was obtained. 
\par
The derivation \cite{yeu} of YRD bound required an integration over $k$ involving the 
structure factors at $t$ and $t_w$, viz., 
\begin{equation}\label{e_yrd}
C_{\textrm{ag}} \sim \ell^{-\lambda} \leq \ell^{d/2}\int_{0}^{b}{dk k^{d-1} [{S(k,t_{w})} {\tilde{S}(k\ell)}]^{1/2}},
\end{equation}
where $\tilde{S}$ is a time independent scaling function and 
$b=\frac{2\pi}{\ell}$, $\ell$ being the domain length at time $t$. 
The bound in Eq. (\ref{e_yrdbound}) follows when $S(k,t_{w})$ in Eq. (\ref{e_yrd}) is replaced 
by its small $k$ behavior as in Eq. (\ref{e_smallk}).
Given that the growth in the considered case is slow, 
value of $b$, i.e, the range of integration in 
$k$-space is bigger, at a given time, compared to higher temperature scenario where the growth is faster. 
This provides further larger ``effective'' value of $\beta$ in this case. 
Nevertheless, question remains, as mentioned above, with the increase of the system 
size a constant value of $S(k,t_{w})$ may be seen for $T=0$ and at very late time the value of $b$  
may fall in that region. This will raise the value of the lower bound. In that case 
do we expect a crossover in the value of $\lambda$, from the value mentioned above, 
to the LM one? This will certainly be interesting to check. However, for that purpose, 
system sizes much larger than the ones considered here must be run for extremely long 
time. This exercise is not within our ability at the moment, 
given the limitation of computational resources 
available to us. Nevertheless, the system size $L=512$ studied in this work 
is extremely large and contains more than $0.13$ billion spins. To our 
knowledge, there exists only one study in the literature that considered comparable 
system size \cite{cor}. 
\section{Conclusion}
We have studied aging property \cite{zan} during ordering in the 3D Ising model without conservation 
of order-parameter. Monte Carlo simulation \cite{lan} results  
for quenches from infinite to zero temperature are presented for the two-time autocorrelation 
function. It has been shown, like in high temperature case \cite{mid}, the decay of this 
correlation function, as a function of $x=\ell/\ell_{w}$, is a power-law, with an exponential 
correction for small $x$. The exponent for the power-law, however, is much smaller 
than that for the high temperature decay \cite{mid}. While in the high $T$ case the exponent is consistent 
with a theoretical prediction by Liu and Mazenko \cite{liu}, that obeys a lower bound provided 
by Fisher and Huse (FH) \cite{dfish}, for $T=0$ this lower bound is violated. This is an extremely 
striking observation. Furthermore, the pattern at $T=0$ is different from that 
of the high temperature. For $T=0$, the Ohta-Jasnow-Kawasaki function \cite{oh} does 
not describe the two-point equal-time correlation function well. 
The origin of this difference has been discussed. 
We argue, this deviation is responsible for the violation of the FH bound. In 
fact our result shows that the aging exponent obeys another bound, obtained by 
Yeung, Rao and Desai \cite{yeu}, that can account for the structural anomaly mentioned above. 
\par
In this work the results from $T=0$ are compared with those from $0.6T_c$ that lies above 
the roughening transition temperature, $T_{R}$. In future we will perform more 
systematic study by gradually varying $T$. This will provide information on whether the surprising 
features that have been observed are only a zero temperature property or there is a 
gradual cross-over from $T=0$ to a higher temperature. This will also reveal 
if $T_{R}$, related to the interface broadening, is responsible for the unusual structure and dynamics.
 
~${*}$ das@jncasr.ac.in

\end{document}